# Single-molecule mid-IR detection through vibrationally-assisted luminescence

Rohit Chikkaraddy*[1], Rakesh Arul[1], Lukas A. Jakob[1], and Jeremy J. Baumberg*[1]

[1] NanoPhotonics Centre, Cavendish Laboratory, Department of Physics, JJ Thompson Avenue, University of Cambridge, Cambridge, CB3 0HE, United Kingdom

**Abstract**

**Room temperature detection of molecular vibrations in the mid-infrared (MIR, $\lambda$=3-30μm) has numerous applications including real-time gas sensing, chemical reactivity, medical imaging, astronomical surveys, and quantum communication[1,2]. However, MIR detection is severely hindered by thermal noise, hence current technologies rely on energy-intensive cooled semiconductor detectors (mercury cadmium telluride, MCT)[3–5]. One way to overcome this challenge is to upconvert the low-energy MIR light into high-energy visible wavelengths ($\lambda$=500-800nm) where detection of single photons is easily achieved using silicon technologies[6,7]. This process suffers from weak cross sections and the mismatch between MIR and visible wavelengths, limiting its efficiency. Here, we exploit molecular emitters possessing both MIR and visible transitions from molecular vibrations and electronic states, coupled through Frank–Condon factors. By assembling molecules into a nanoscale cavity and continuously optically pumping them below the electronic absorption band, we show the transduction of MIR light absorbed by the molecular vibrations. The upconverted signal is observed as enhanced high-energy luminescence. Combining Purcell-enhanced visible luminescence with enhanced rates of vibrational pumping gives transduction efficiencies exceeding 10%. By down-scaling the cavity volume below 1nm$^3$, we show MIR detection of single-molecular bonds, inaccessible to any previous detector.**

Atomic bonds possess MIR resonances corresponding to vibrational transitions between states. While these are thermally populated at room temperature, electronic states with transition energies in the visible have negligible thermal occupation. Detection of visible light is efficient down to the single photon level, but detection of MIR photons remains challenging. However, these states can interact together. Quantum systems in which such vibrations are coupled to electronic transitions have proved fertile for quantum manipulation, for instance using trapped ions which are entangled through their vibrational motions in a trap[8]. More recent efforts to create solid state quantum implementations have exploited quantum dots or defect centres embedded in micro-mechanical resonators as optomechanical elements[9,10]. Here we demonstrate their ultimate scaling-down to the sub-nm scale, by coupling electronic and vibrational transitions in molecular bonds, down to single dye molecules. We show that this provides a practical way to detect MIR light which is normally hindered by thermal noise[11], through upconverting the low energy photons to visible wavelengths.

Demonstrations of MIR upconversion detection predominantly use parametric mixing in non-linear crystals[6], interferometric upconversion with entangled photon pairs[12], or non-degenerate two-photon absorption[13]. These methods require high-intensity ultrafast pulsed lasers to drive the nonlinear susceptibilities. In optomechanical cavity-assisted wave-mixing, MIR detection is possible through specific vibrational bonds which are both infrared and Raman active[14,15], but the Raman process is intrinsically low cross-section.

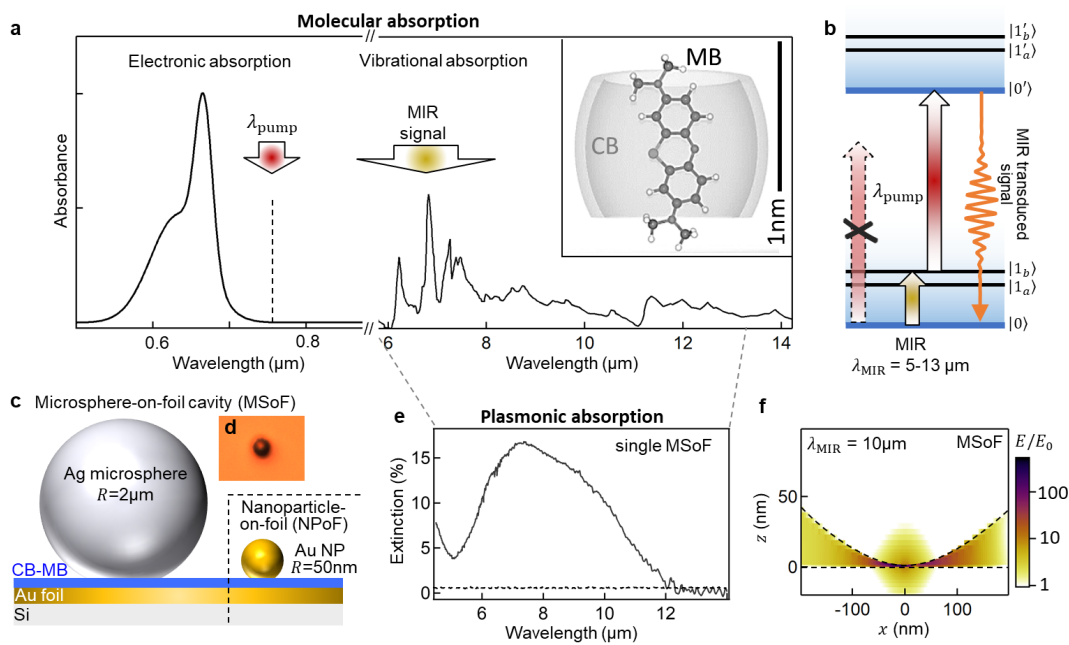

**Figure 1: MIR vibrationally-assisted luminescence (MIR-VAL).** (**a**) Electronic and vibrational absorption spectra of methylene blue (MB) molecule in solution. Arrows indicate NIR and MIR tuning. Inset shows the supramolecular assembly of MB inside a cucurbit[7]uril (CB) host molecule. (**b**) Energy diagram of electronic and vibrational levels of MB molecule, indicating the NIR pump (red arrow) drives the electronic transition only in the presence of MIR (yellow arrow). (**c**) Silver-coated microsphere constructed on a thin metal foil (MSoF) along with alternative gold nanoparticle-on-foil (NPoF) construct. (**d**) Bright field optical image of an individual MSoF construct. (**e**) Fourier transform infrared spectroscopy of individual MSoF

cavity (solid grey), compared to simulated extinction of a NPoF cavity (dotted grey). (**f**) Simulated near-field of MSoF cavity at $\lambda_{\text{MIR}}$=10μm.

Here we show MIR vibrationally-assisted luminescence (MIR-VAL) from molecular emitters. This scheme exploits the coupling between visible electronic transitions and the MIR vibrational transitions in a molecule. The electronic and vibrational absorption spectra of methylene blue (MB) molecules (Fig.1a) show characteristics of a typical dye molecule. The strong dipolar electronic transition of MB has dipole moment 3.8D at 665nm corresponding to $|0\rangle \rightarrow |0'\rangle$ from ground state to first singlet absorption[16]. The vibronically-mediated absorption $|0\rangle \rightarrow |1'\rangle$ appears as a shoulder peak at 620nm with strength governed by Franck-Condon coupling factors. When MB absorbs wavelengths shorter than 700nm, the electronically-excited molecule relaxes to the ground state via down-shifted photoluminescence (PL) emission at 680nm ($|0'\rangle \rightarrow |0\rangle$) and 710nm ($|0'\rangle \rightarrow |1\rangle$) following Kasha's rule[17].

In our MIR detection scheme, the molecule is now primed with continuous-wave (CW) pumping ($\lambda_{\text{pump}} \geq$750nm) at energies below the MB absorption band, so no PL is observed (Fig.1b). In the presence of MIR light, specific vibrational levels in each molecule become populated ($|0\rangle \rightarrow |1_a\rangle, |1_b\rangle$) while remaining in the ground electronic state. Here $|1_a\rangle, |1_b\rangle$ are different bond vibrations $a, b$ within the MB molecule. Once the vibrational states are populated, the absorption of the pump becomes allowed and drives the molecule from ground- to electronically-excited state. This transition now relaxes to the ground state giving PL at wavelengths shorter than the pump, resulting in anti-Stokes photoluminescence (aS-PL, on the higher energy anti-Stokes side of the pump), forming the MIR transduced signal. Such vibrationally-assisted PL is analogous to optical spin read-out methods recently developed to record nuclear magnetic resonance of quantum systems at the nanoscale[18,19].

At room temperature, vibrational decays ($10^{-12}$ s) are thousand-fold faster than electronic decays ($10^{-9}$ s), limiting the normal transduction efficiency of isolated molecules to <$10^{-7}$ for CW pumping[7]. While aS-PL can be observed for intense pulsed excitation[20,21] such requirements are not practical for compact MIR detectors. To overcome this, here the molecules are placed inside a deep-subwavelength optical cavity formed of nanogap-spaced metal walls which can provide million-fold up-conversion enhancements. The rate of light emission is Purcell enhanced ($P \propto 1/V$) where $V$ is the cavity mode volume, while the rate of excitation is also amplified by the optical near-field intensity ($E^2 \propto 1/V$) at the location of the molecule, for field $E$ oriented along its transition dipole. An ideal optical cavity will thus have the smallest mode volume with strong overlapping near-field intensities at both pump and MIR wavelengths.

To construct such small nanocavities with double resonances tuned to visible/MIR and containing molecules precisely oriented along the optical field, we use bottom-up nano-assembly. Monodisperse silver-coated glass microspheres (AgMS) are drop-cast onto a thin (10nm) foil of flat Au with molecular emitters sandwiched in the gap formed between them

(Fig.1c). The resulting microsphere-on-foil (MSoF) system has a gap that is accurately controlled to sub-nm, is straightforward to characterize, easily and repeatably made, and is scalable to large areas[22,23]. The diameter of each AgMS ($2R$=4μm) is tuned to achieve MIR resonance with the intense plasmonic peak centered around 8μm (extinction coefficient >10%, Fig.1e). The interaction between each AgMS and its image in the underlying foil forms an effective dimer with strong electromagnetic enhancements ($E^2/E_0^2 > 10^5$) perpendicular to the surface and ultralow mode volume (Fig.1f).

The molecular dye MB is assembled into the nanogaps using the host-guest chemistry of macrocyclic cucurbit[7]uril (CB) molecules. CB provides a hydrophobic environment which isolates individual MB molecules and improves their photostability[24] (Fig.1a). The carbonyl portals at either end of the 0.9 nm-high CB bind to Au, scaffolding the MB dipole near-perpendicular to the metal surfaces for optimal coupling to the electromagnetic near-field. We compare the MSoF with smaller diameter nanoparticle-on-mirror (NPoF) constructs ($2R$=0.1μm, Fig.1c) which support a NIR resonance but only broad non-resonant MIR response (dotted grey line in Fig.1e). Previous studies that tuned CB-MB fill fractions show that there are, on average, $\bar{n}$=40 MB molecules in the gap for 1% of CB filled with MB, after taking into account the different sizes of nanofacets[24]. No PL is observed when the MSoF gaps are empty of MB dyes.

Assembling MSoF nanocavities on thin (200 μm) silicon substrates allows MIR light to couple from the underside while simultaneously visible pumping from the top side (Fig.2a). For simultaneous control measurements and improved signal to noise, both MIR and pump are modulated at 300 kHz. The emitted light from the sample is collected from the top side and routed to a monochromator-mounted silicon detector array after rejecting the pump beam with dielectric filters. The observed spectrum to lower (Stokes) energy shows surface-enhanced resonant Raman scattering (SERRS) peaks of MB molecules assembled within the gap (Fig.2b). In the absence of MIR light, the decay of intensity on the anti-Stokes (higher energy) side is attributed to electronic Raman scattering (ERS) from the metal[25]. The SERRS peaks are observed only when the pump is focused directly onto the MSoF and not on the flat Au. In the presence of MIR light at 1623cm$^{-1}$ the Stokes is unchanged, but on the anti-Stokes side we observe an upconverted PL with characteristic peak at 680nm corresponding to MB emission. For $\lambda_{\text{pump}}$ = 750nm the effective wavelength of summed photon energies $1/\lambda = 1/\lambda_{\text{pump}} + 1/\lambda_{MIR}$ = 669 nm excites the $|0\rangle \to |0'\rangle$ transition of MB. The MIR transducted aS-PL intensity is increased >180% above the background (Fig.2c) compared to without MIR light (Fig.2d).

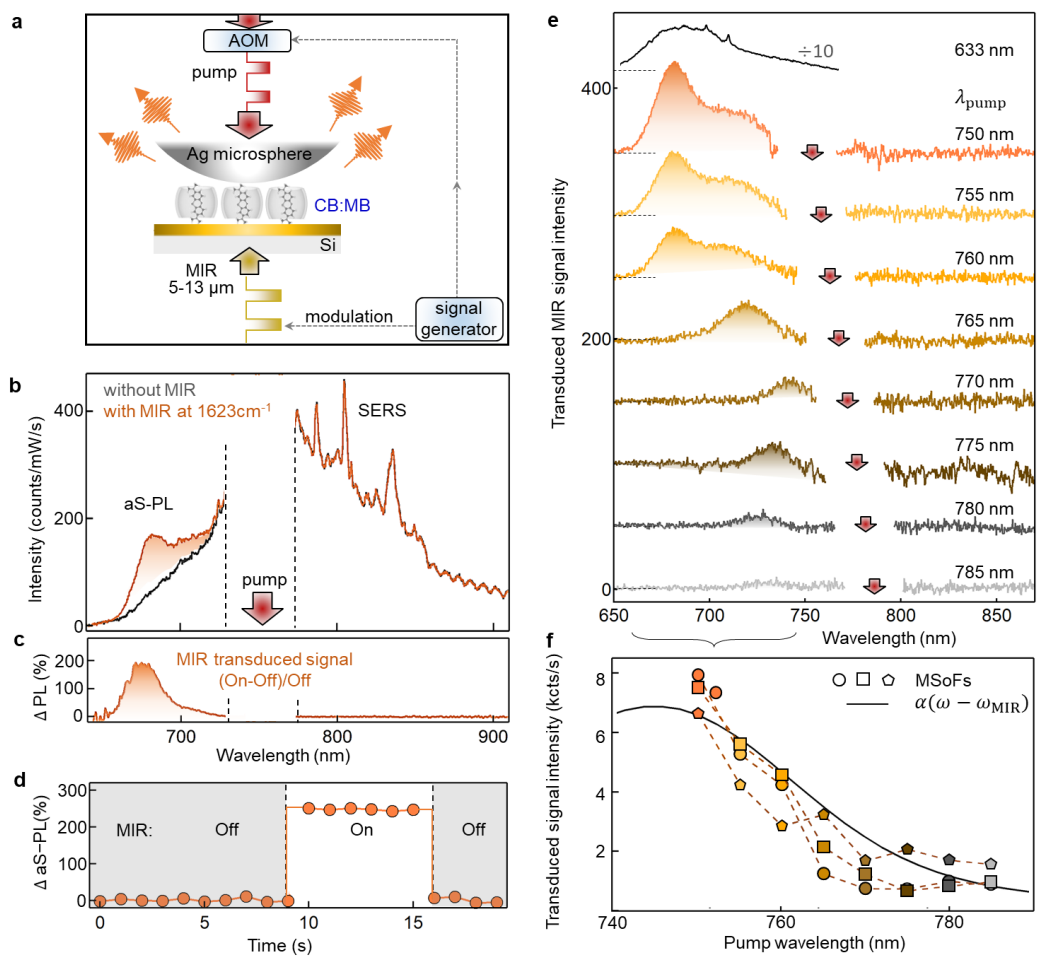

**Figure 2: Transducted MIR in vibrationally-assisted luminescence.** (**a**) MIR detection setup with pump illuminating an individual MSoF cavity. Incident light modulated (dashed lines) using acousto-optic modulator (AOM) for pump, QCL laser for MIR. (**b**) Stokes and anti-Stokes emission from pump, with (orange) and without (grey) MIR light of average power 0.5µW/µm², averaged over 10 scans on a single MSoF cavity. (**c**) Percentage difference in emitted signals transducted by MIR in (b). (**d**) Modulation of aS-PL intensity switching the MIR beam on/off. (**e**) MIR transducted signal when tuning the pump from $\lambda_{\text{pump}}$ = 750nm (top) to 785nm (bottom) in 5 nm steps (arrows indicate $\lambda_{\text{pump}}$). Direct visible pumping at 633nm shown for comparison (top, grey curve). (**f**) Spectrally-integrated transducted signal from (e) vs $\lambda_{\text{pump}}$ for three different MSoF cavities, compared with absorption spectra of MB molecules in solution, rigidly-shifted by $\omega_{MIR}$ (solid line).

As the pump wavelength is tuned from 750nm to 785nm (Fig.2e) while keeping the MIR wavelength at 6.16µm (1623cm⁻¹), the observed vibrationally-assisted luminescence signal decreases. This is because the summed photon energy now drops below the $|0\rangle \rightarrow |0'\rangle$ transition of MB. There is no change in the SERRS spectra on the Stokes side, confirming that MB dyes in the gap are not damaged. The spectrally-integrated transducted signal *vs* pump wavelength follows the shape of the absorption curve of MB in solution (Fig.2f), when rigidly shifted to lower energy by the MIR photon energy. This result is consistent for each MSoF nanocavity measured, confirming that the observed aS-PL is the direct effect of MIR transduction. The signal is strong enough to be now seen at room temperature and for CW pumping. Intriguing changes in aS-PL spectra arise for $\lambda_{\text{pump}} > 765$nm, corresponding to upconverted excitation below the $|0\rangle \rightarrow |0'\rangle$ energy. These changes depend on the precise MIR plasmon resonance in each MSoF (Supplementary information, Fig.S1 and Fig.S2), thus implicating plasmon enhancements. Exciting the vibrations by MIR light induces nuclear displacements set by the Huang-Rhys factors[26], controlling the overall transfer of energy from $|1_b\rangle$ to $|0'\rangle$. Here the MIR-VAL intensity using 0.5µW/µm² MIR (our highest MIR power) is >10% of that when directly exciting $|0\rangle \rightarrow |0'\rangle$ with a 633nm pump of the same pump power as used in MIR-VAL.

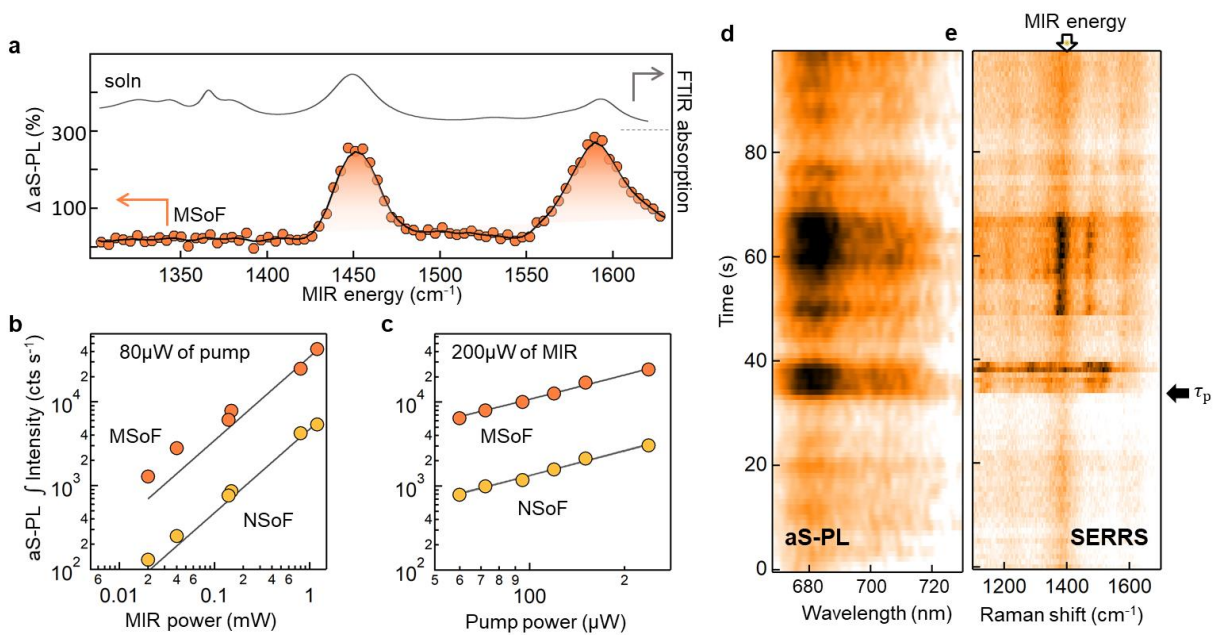

**Figure 3: MIR detection efficiency and transducted single-molecule absorption.** (**a**) aS-PL intensity when tuning MIR photon energy (orange points) compared to FTIR measurements of MB in solution (grey). (**b,c**) Power-dependent integrated aS-PL intensity for increasing (b) MIR and (c) pump average powers, using MSoF (orange) and NPoF (yellow). Detection noise floor is 100 cts/s. (**d,e**) Repeated aS-PL and SERRS spectra for single MSoF cavity with 0.5mW/µm² pump at 750nm, top arrow shows MIR tuning. The transient picocavity event starts at $\tau_P$.

To confirm that the MIR transducted signals originate from vibrations of the MB molecule, we tune the MIR energy from 1300 to 1630cm⁻¹ keeping the pump fixed at 750nm and using an individual MSoF cavity (Fig.3a). The percentage change in aS-PL intensity closely follows the vibrational absorption spectrum of MB molecules in solution obtained from conventional FTIR spectroscopy. Subtle differences are attributed to the effect of the plasmonic nano-gap environment, compared to the normal ensemble solution spectra. The MIR transducted signal from $\bar{n}$=40 molecules has >10:1 signal to noise with spectral resolution limited only by the linewidth of the quantum cascade laser (QCL) and step size used (4.2cm$^{-1}$). This is impossible to achieve in conventional FTIR measurements, even through surface-enhanced infrared absorption (SEIRA[30]) with single MSoF cavities (Fig.1e).

To demonstrate the generality of this aS-PL scheme, MSoF systems are similarly assembled with rhodamine B molecules in the gap (Supplementary information, Fig.S3). The MSoF cavity is now pumped with $\lambda_{\text{pump}} = 663$nm, below the $|0\rangle \rightarrow |0'\rangle$ transition energy of rhodamine B. An even larger aS-PL intensity change is now observed of >40% for 0.5μW/μm$^2$. MIR tuning experiments again confirm that the vibrational spectrum of rhodamine B controls the transduction, and again is subtly different from ensemble measurements (Supplementary information, Fig.S4).

The current MIR detection limit is quantified through the observed power dependence (Fig.3b,c). MIR transducted intensities are linear in both MIR and pump intensities, and linear with molecular concentration ($\bar{n}$) in the gap (Supplementary information, Fig.S5). MSoF cavities provide ten-fold better MIR-VAL efficiencies compared to NPoF cavities. This clearly demonstrates how MIR resonances enhance the coupling of MIR light to the vibrational transitions. The system can be suitably modelled with rate equations of a four-level system (Supplementary information, Fig.S6). While the Purcell-enhanced aS-PL from $|0'\rangle \rightarrow |0\rangle$ is incoherent (due to relaxation by intramolecular rotations and vibrations), the $|0\rangle \rightarrow |0'\rangle$ transition is coherently driven by the combination of MIR and pump light.

From the noise floor of 100 dark counts/s (which drops significantly when spectrally-integrating onto silicon detectors) we easily measure average MIR powers of 2.5nW/μm$^2$ using 80μW/μm$^2$ of pump for the MSoF cavities. This corresponds to a noise equivalent power (NEP) of 31nW/Hz$^{1/2}$ which, accounting for the <0.1% MIR coupling efficiency, implies achievable NEP below 30pW/Hz$^{1/2}$, better than available thermopiles and close to state-of-the-art MCT cooled detectors. One source of noise is set by fluctuations in the ERS background emission due to the unstable gold nanostructure within the gap, and can be suppressed by increasing the separation of $\lambda_{\text{PL}}$ and $\lambda_{\text{pump}}$. Additional opportunities are to explore upconverted MIR from C-H bond vibrations at $\lambda_{\text{MIR}}$=3.3μm or overtones of water in the range $\lambda_{\text{MIR}}$=1-2μm.

The aS-PL transduction of MIR light has a unique advantage in the extremely fast rise time of detected PL. Using time-correlated single-photon lock-in methods[27], this PL rise time is resolved < 100ns (Supplementary information, Fig.S5) which is currently limited by the QCL source rise time. Theoretically, since the aS-PL is Purcell-accelerated, high pump rates should give <10ps response implying >100GHz modulation speeds. A particular advantage is thus high frequency lock-in detection, to further reduce the noise floor. This stimulates the need for developing matched low-cost high-bandwidth MIR sources[28].

As a final demonstration that this MIR detector can observe single molecules, the pump is increased to 0.5mW/μm$^2$. At this power, the light inside the nanocavity acts to transiently pull single metal atom protrusions out of the facets. These generate a picocavity[29] where light is extra confined around the adatom to volumes below 1nm$^3$, and focused onto a single-molecule. Such transient events are observed as new vibrational lines in SERRS (Fig.3d,e after $\tau_P$), whose spectral fluctuations evidence their single-molecule origin[29]. The MIR energy is now tuned to one of these new vibrations (top arrow Fig.3e), and both the aS-PL and SERRS are simultaneously recorded. The aS-PL shows transient variations in intensity which exactly correlate in time with the picocavity events in SERRS, and give >200-fold PL enhancement over all the other dye molecules in the nanocavity outside this picocavity. This therefore evidences the first experimental evidence for MIR absorption of a single vibrational bond (SEIRA). Ongoing efforts to stabilize these picocavities will thus allow $\lambda_{\text{MIR}}$-swept MIR-VAL to extract full SEIRA spectra of single molecules, a long-term goal for plasmonics.

In summary, we demonstrate a unique way to optically read-out molecular vibrations down to the single-molecule level. This method unambiguously shows vibration-mediated upconversion of MIR to visible light through antiStokes photoluminescence. The large detector bandwidth (>GHz), room temperature (or other) operation, straightforward large-scale fabrication, and broad MIR tuneability with potential for THz detection make this a highly promising scheme. Future experiments will explore MIR multiplexing, capability of using many types of dye molecules as well as solid-state emitters (such as PbS quantum dots), alternate types of MIR resonators, and LED pumping. This demonstration opens unique possibilities not just for molecular spectroscopy and sensing but has wide implications in quantum preparation of vibrational states for mode-selective chemistry and nano-optics.

**Materials and Methods**

**Sample preparation.** To prepare the thin mirror, we deposit 10nm of Au onto clean Si membranes of 200μm thickness with a deposition rate of 0.5 Å/s (Moorfield nanoPVD-T15A thermal evaporator). The Au-coated substrates are dipped into 0.1 mM MB-CB solution for 12h resulting in a self-assembled monolayer of MB-CB on the gold. For Ag MSoF optical cavities, 4μm diameter silver-coated microspheres (Cospheric) dispersed in ethanol are deposited directly onto the MB-CB assembled Au-coated Si substrates. The deposition time is kept below 90s, resulting in well-dispersed AgMSs. Lastly, the samples are rinsed thoroughly with ethanol to remove any unbound AgMSs.

**Infrared microscopy.** FTIR reflection spectra on individual MSoF cavities are recorded using a Shimadzu AIM-9000 FTIR microscope, equipped with liquid-nitrogen-cooled MCT detector and Cassegrain 15X objective with 0.5 NA. The spectra are referenced to a clean gold mirror.

**Experimental setup.** All SERS and aS-PL spectroscopy measurements are performed in a custom-built dual-channel microscope. A tunable CW TiS laser is used as the pump, which is spectrally filtered using an angle-tuned 785nm cleanup filter. The laser is focused into the sample with a 0.9NA objective lens with 150μW/μm$^2$ average power on the sample (where not otherwise noted). The back-scattered light is collected from the sample objective and is filtered with two angle-tuned notch filters before routing it to a Shamrock i303 spectrograph and a Newton EMCCD (Andor). For imaging, the reflected light collected through the same objective lens is directed to a camera (Lumenera Infinity3-1).

For the MIR light source, a LaserTune IR source (Block) with wavelength range of 5.4 - 13μm is used (1635 - 780 cm$^{-1}$) with a maximum average output of 500 μW (~2x4 mm beam, collimated) with <5% duty cycle. The pump (MIR light) is co-aligned with the probe (visible light) using a 0.5 NA Cassegrain objective lens. The modulation of pump laser is synced with the QCL using an acousto-optical-modulator and function generator. The sample is placed on a fully automated motorized stage (Prior Scientific H101) which is controlled with code written in Python.

For single-photon time-correlated measurements, arrival times of all photons at the detector (Micro Photon Devices PDM PD-100-CTD) and reference signals (MIR laser trigger) are continuously recorded by time-to-digital converters on a field-programmable gate array board. Comparing the photon timestamps with the reference signal allows recreating the periodic perturbation of the SERS signal by the MIR laser in time, integrated over millions of modulation cycles. This single-photon lock-in detection scheme is described in more detail elsewhere[27].

**Acknowledgements**

We acknowledge support from European Research Council (ERC) under Horizon 2020 research and innovation programme PICOFORCE (Grant Agreement No. 883703), THOR (Grant Agreement No. 829067) and POSEIDON (Grant Agreement No. 861950). We acknowledge funding from the EPSRC (Cambridge NanoDTC EP/L015978/1, EP/L027151/1, EP/S022953/1, EP/P029426/1, and EP/R020965/1). R.C. acknowledges support from Trinity College, University of Cambridge. R.A. acknowledges support from the Rutherford Foundation of the Royal Society Te Apārangi of New Zealand, and the Winton Programme for the Physics of Sustainability.

**Author Information**

**Corresponding Author**

* Dr Rohit Chikkaraddy, rc621@cam.ac.uk
* Prof Jeremy J Baumberg, jjb12@cam.ac.uk

**Author Contributions**

R.C and J.J.B conceived and designed the experiments. R.C. performed the experiments with input from R.A and L.A.J. R.A. carried out the FTIR and RhB measurements. R.C performed the simulation and the analytical modelling. R.C. and J.J.B. analysed the data. R.C. and J.J.B. wrote the manuscript with input from all authors.

**Conflict of interest**

The authors declare no competing financial interest, and that they have filed IP on this technology

**Supporting Information**. A Supporting Information document is also provided, with additional images and information. Source data can be found at https:// (to be provided on acceptance).